\documentclass[twoside]{dis07}
\usepackage[latin1]{inputenc}
\usepackage[dvips]{graphicx,epsfig,color}
\usepackage{wrapfig,rotating}
\usepackage{amssymb,amsmath,array}

\begin{document}
\title{Project to install roman pot detectors at 220 m in ATLAS}

%***********************************************************************
% AUTHORS INFORMATION AREA
%***********************************************************************
\author{Christophe Royon
%
% Optional short acknowledgment: remove next line if non-needed
\thanks{On behalf of the RP220 Collaboration}
%
% DO NOT MODIFY THE FOLLOWING '\vspace' ARGUMENT
\vspace{.3cm}\\
%
% Addresses and institutions (remove "1- " in case of a single institution)
DAPNIA/Service de physique des particules, \\ CEA/Saclay, 91191 
Gif-sur-Yvette cedex
%
% Remove the next three lines in case of a single institution
}
%***********************************************************************
% END OF AUTHORS INFORMATION AREA
%***********************************************************************

\maketitle

\begin{abstract}
We give a short description of the project to install roman pot detectors at 220
m from the interaction point in ATLAS. This project is dedicated to hard  diffractive
measurements at high luminosity.
\end{abstract}

\section{Introduction}
The motivation to install roman pot detectors at 220 m within ATLAS is quite
clear. It extends nicely the project of measuring the total cross sections
using roman pots at 240 m \cite{alfa} by measuring hard diffraction at high
luminosity in ATLAS in the LHC. As we will see in the following, it is also
complementary to the FP420 project which aims at tagging protons at 420 m.

The physics motivation of this project corresponds to different domains of
diffraction:
\begin{itemize}
\vspace{-0.3cm}
\item A better understanding of the inclusive diffraction mechanism at the LHC by studying
in detail the structure of pomeron in terms of quarks and gluons as it was done
at HERA \cite{h1zeus}. Of great importance is also the measurement of the exclusive production
of diffractive events \cite{olda} and its cross
section in the jet channel as a function of jet transverse momentum. Its
understanding is necessary to control the background to Higgs signal.
\vspace{-0.3cm}
\item Looking for Higgs boson diffractive production in double pomeron exchange in
the Standard Model or supersymmetric extensions of the Standard Model \cite{us}. This is
clearly a challenging topic especially at low Higgs boson masses where the
Higgs boson decays in $b \bar{b}$ and the standard non-diffractive search is
possible. We will detail in the following the trigger strategy. 
\vspace{-0.3cm}
\item Sensitivity to the anomalous coupling of the photon by measuring the QED
production cross section of $W$ boson pairs. This might be the best way to
access the anomalous coupling before the start of the ILC.
\vspace{-0.3cm}
\item Photoproduction of jets
\vspace{-0.3cm}
\item Other topics such as looking for stop events or measuring the top mass
using the threshold scan method \cite{stephane} which will depend strongly on
the production cross section.
\end{itemize}

\section{Roman pot design and location}
We propose to install roman pots in ATLAS at 216 and 224 m on each side of the
main ATLAS detectors. The project is a collaboration between the physics
institutes and universities of Prague, Cracow, Stony Brook, Michigan State
University, LPNHE (Paris 6), Giessen, and in addition the University of Chicago
and the Argonne National Laboratory for the timing detectors.

The roman pot design follows as close as possible the design which is currently
used by the TOTEM collaboration and the Luminosity group of the ATLAS
collaboration which aims at measuring the total cross section using roman pots
at 240 m. The only difference is that we only need the horizontal arms and not
the vertical arms since hard diffractive protons are scattered horizontally.
We will follow the TOTEM experience to build the roman pots in Vakuum Praha and
to use the same technics for the step motors and the LVDT system.

\begin{wrapfigure}{r}{0.5\columnwidth}
\centerline{\includegraphics[width=0.4\columnwidth]{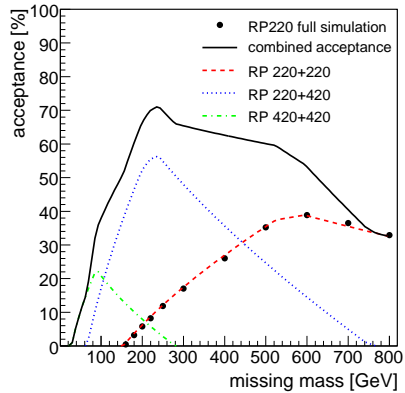}}
\caption{Roman pot detector acceptance as a function of missing mass
assuming a 10$\sigma$ operating positions, a dead edge for the detector of 50
$\mu m$ and a thin window of 200 $\mu m$.}\label{Fig1}
\end{wrapfigure}

Assuming one can go  down to 10 (resp. 15) $\sigma$ from the beam center, it is
possible to measure protons with $\xi>0.01$, and $\xi>0.012$ on each side of
ATLAS (resp. $\xi >0.014$, $\xi >0.016$) where $\xi$ is the momentum fraction of
the initial proton carried away by the Pomeron \cite{note}. This can be translated
in missing mass acceptance as illustrated in Fig 1. The missing mass acceptance using only
the 220 m pots starts at 135 GeV, but increases slowly as a function of missing
mass. It is clear that one needs both FP420 and RP220 projects, or in other
words the possibility to detect scattered protons at 220 and 420 m to obtain a good
acceptance on a wide range of masses since most events are asymmetric (one tag
at 220 m and another one at 420 m). The precision on mass reconstruction using
either two tags at 220 m or one tag at 220 m and another one at 420 m is of the
order of 2-3 \% on the full mass range. This shows the advantage of this
measurement which allows to give a very good mass resolution on a wide range of
masses, and thus to detect Higgs bosons at low masses decaying into $b \bar{b}$.
The idea is to enhance the signal over background ratio by benefitting from the
good resolution of the detectors and the suppression of the $b$ jet background due
to the $J_z=0$ suppression rule for $b$ jet exclusive production.

\section{Detector inside roman pots}
We propose to put inside the roman pots two kinds of detectors, namely Silicon
detectors to measure precisely the position of the diffracted protons, and the mass
of the produced object, such as the Higgs boson, and
$\xi$, and precise timing detectors.

The position detectors will consist in either five layers of Silicon strips of
50 $\mu$m and two additional layers used for triggering, or 3D Silicon
detectors if they are available industrially by the time we need to instal the 
roman pots. If the Silicon strip option is chosen, there will be four different
orientations, namely X, Y, U, and V (U and V being orientated within 45 degrees
with respect to X and Y). The strip size will be 50 $\mu$m and the detector size
about 2 cm, which allows a measurement up to $\xi \sim 0.15$. The Silicon strip
detectors will be edgeless which means that the dead edge will be of the order
of 30-50 $\mu$m so that we can move the detector as close to the beam as
possible without losing some acceptance due to the dead edge. The detectors will
be read out by the standard ABCNext chip being developped in Cracow for the
Silicon detector of ATLAS. The latency time of the ABCNext chip is of the order
of 3.5 $\mu$s which gives enough time to send back the local L1 decision from
the roman pots to ATLAS (see the next paragraph about trigger for more detail),
and to receive the L1 decision from ATLAS, which means a distance of about 440
m. It is also foressen to perform a slight modification of the ABCNext chip to
include the trigger possibilities into the chip.
The other option is to use 3D Silicon detectors using the same readout system
as before (ABCNext chip).  These detectors 
use a lateral electric field, instead of vertical in conventional planar 
techniques. Holes of the order of 10 $\mu$m crossing the 
full thickness of the 
detector  are filled with a conductive medium in order to collect 
the ionisation (electrons or holes) depending on the applied bias. Both kinds of
options will be tested in Prague and in Saclay using the full electronics chain
(including the ABCNext chip) and a laser or a radioactive source. Beam tests at
DESY or CERN are also foreseen.
It is planed to install the roman pot together with the Silicon detectors during
a shut down of the LHC in 2009-2010.

The timing detectors are necessary at the highest luminosity of the LHC to
identify from which vertex the protons are coming from. It is expected that up
to 35 interactions occur at the same bunch crossing and we need to identify from
which interaction, or from which vertex the protons are coming from. A precision
of the order of a few mm or 5-10 ps is required to distinguish between the
different vertices and to make sure that the diffracted protons come from the
hard interactions. Picosecond timing detectors are still a challenge and are
developped in a collaboration between Saclay, Stony Brook,
the University of Chicago and Argonne
National Laboratory for medical and particle physics applications. The proton
timings will be measured in a crystal of about 2.5 cm located inside the roman
pots, and the signal will be read out by Micro-Channel Plates Photomultipliers
developped by Photonis. The space resolution of those detectors should be of the
order of a few mm since at most two protons will be detected in those detectors
for one given bunch crossing at the highest luminosity. The detectors are read
out with a Constant Fraction Discriminator which allows to improve the timing
resolution significantly compared to usual electronics.
A first version of the
timing detectors is expected to be ready in 2009-2010 with a worse resolution of
40-50 ps, and the final version by 2012 with a resolution of 5-10 ps.

\section{Trigger principle and rate}
In this section, we would like to give the principle of the trigger using the
roman pots at 220 m as well as the rates obtained using a simulation of the
ATLAS detector and trigger framework.

The principle of the trigger is shown in Fig. 2 in the case of a Higgs boson
decaying into $b \bar{b}$ as an example.
The first level trigger comes directly from two different Silicon strip layers
in each roman pot detector. It is more practical to use two dedicated planes for
triggering only since it allows to use different signal thresholds for trigger
and readout. The idea is to send at most five strip addresses which are hit
at level 1. A local trigger is defined at the roman pot level on each side of
the ATLAS experiment by combining the two trigger planes in each roman pot and
the roman pots as well. If the hits are found to be compatible (not issued by
noise but by real protons), the strip addresses are sent to ATLAS, which allows
to compute the $\xi$ of each proton, and the diffractive mass. This information
is then combined with the information coming from the central ATLAS detector,
requesting for instance two jets above 40 GeV in the case shown in Fig. 2. At
L2, the information coming from the timing detectors for each diffracted proton
can be used and combined with the position of the main vertex of ATLAS to check
for compatibility. Once a positive ATLAS trigger decision is taken (even without
any diffracted proton), the readout informations coming from the 
roman pot detectors are sent to ATLAS as any subdetector.

The different trigger possibilities for the roman pots are given below:
\vspace{-0.3cm}
\begin{itemize}
\item{\bf Trigger on DPE events at 220 m:}
This is the easiest situation since two protons can be requested at Level 1 
at 220 m. Three different options are considered:
\newline
%\begin{itemize}
{\bf - trigger on high mass Higgs} ($M>160$ GeV) given by ATLAS directly
(decay in $WW$, $ZZ$),
\newline
{\bf - inclusive trigger on high mass object} by requesting two high $p_T$ jets
and two positive tags in roman pots,
\newline 
{\bf - trigger on jets} (high $p_T$ jets given directly by ATLAS, and low $p_T$
jet special trigger for QCD studies highly prescaled).
\newline
%\end{itemize}
This configuration will not rise any problem concerning the L1 rate since most
of the events will be triggered by ATLAS anyway, and the special diffractive
triggers will be for QCD measurements and can be highly prescaled.
\vspace{-0.3cm}

\item {\bf Trigger on DPE events at 220 and 420 m}
This is the most delicate scenario since the information from the 420 m pots cannot
be included at L1. The strategy is the following (see Table 1):
%\begin{itemize}
\newline
{\bf - trigger on heavy objects} (Higgs...) decaying in $b \bar{b}$ by requesting
a positive tag (one side only) at 220 m with $\xi < 0.05$
(due to the $420\mathrm{m}$ RP acceptance
in  $\xi$, the proton momentum fractional loss in the $220\mathrm{m}$ roman pot
cannot be too high
if the Higgs mass is smaller than $140\,\mathrm{GeV}$) , and topological cuts
on jets such as the exclusiveness of the process
($(E_{jet1}+E_{jet2})/E_{calo}>0.9$, 
$(\eta_1+\eta_2)\cdot\eta_{220} > 0$, where 
$\eta_{1,2}$ are the pseudorapidities of the two L1 jets, and $\eta_{220}$
the pseudorapidity of the proton in the $220\mathrm{m}$ roman pots). 
This trigger can hold without prescales to a
luminosity up to 2.10$^{33}$ cm$^{-2}$s$^{-1}$,
\newline
{\bf - trigger on jets} (single diffraction, or double pomeron exchange) for QCD
studies: can be heavily prescaled,
\newline
{\bf - trigger on $W$, top...} given by ATLAS with lepton triggers.
%\end{itemize}
\newline
Let us note that the rate will be of the order of 1 Hz at L2 by adding a cut on
a presence of a tag in the 420 pots, on timing, and also on the compatibility of
the rapidity of the central object computed using the jets or the protons in
roman pots.
\end{itemize}

\begin{footnotesize}
\begin{table}
\begin{center}
\begin{tabular}{|c|c|c|c|c|c|}
\hline
${\cal L}$ & $n_{pp}$ per & 2-jet  & RP200 
& $\xi < 0.05$  & Jet \cr
$E_T > 40\,\mathrm{GeV}$ & bunch & rate $[\mathrm{kHz}]$ &
reduction & reduction & Prop.  \cr
 & crossing & $[\mathrm{cm}^{-2}\cdot\mathrm{s}^{-1}]$& factor & factor & \cr
\hline
$1\times10^{32}$ & 0.35 & 2.6 & 120 & 300 & 1200 \cr
$1\times10^{33}$ & 3.5 & 26 & 8.9 & 22 & 88 \cr
$2\times10^{33}$ & 7 & 52 & 4.2 & 9.8 & 39.2 \cr
$5\times10^{33}$ & 17.5 & 130 & 1.9 & 3.9 & 15.6 \cr
$1\times10^{34}$ & 35 & 260 & 1.3 & 2.2 & 8.8 \cr
\hline
\end{tabular}
\end{center}
\caption{L1 rates for 2-jet trigger with $E_T > 40\,\mathrm{GeV}$ and
additional reduction factors due to the requirement of triggering on
diffractive proton at $220\,\mathrm{m}$, and also on jet properties.}
\label{t_trigger}
\end{table}
\end{footnotesize}

%\section{Conclusion}
In this short report, we described the main aspects of the project to install
roman pots at 220 m within ATLAS: Silicon detectors, measurement of the proton
timings, and the trigger properties. This project is aimed to be proposed to
ATLAS and the LHCC together with the FP420 one.

\begin{footnotesize}

% ----------------------------------------------------------------------------
\end{footnotesize}

%\begin{wrapfigure}{r}{1.2\columnwidth}
\begin{figure}
\centerline{\includegraphics[width=0.75\columnwidth,
angle=270]{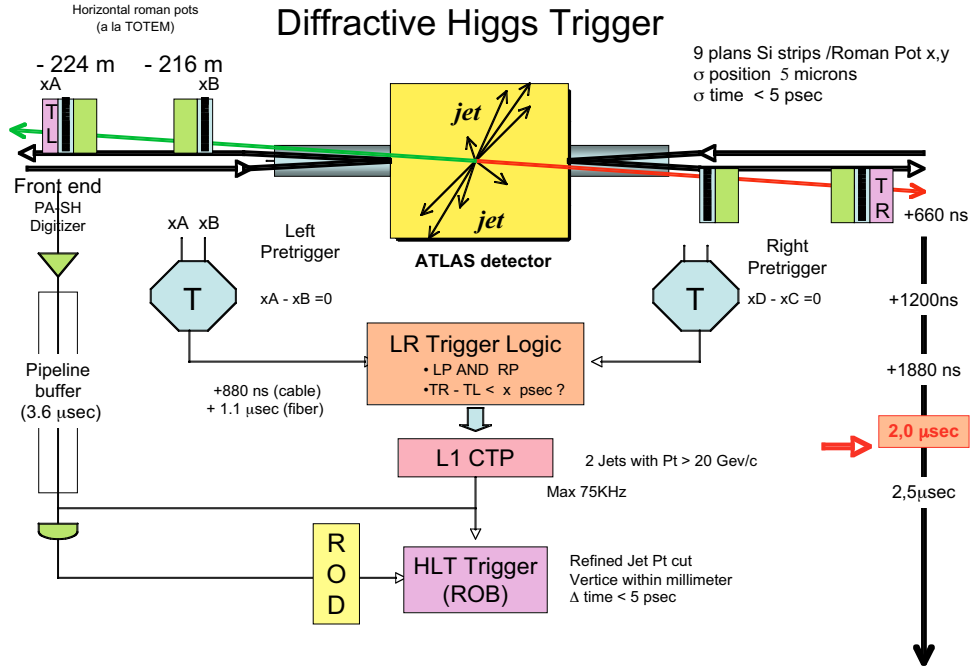}}
\caption{Principle of the L1 trigger using roman pot detectors at 220 m in the
case of a Higgs boson decaying into $b \bar{b}$.}\label{Fig2}
\end{figure}
%\end{wrapfigure}

% ****************************************************************************
% END OF BIBLIOGRAPHY AREA
% ****************************************************************************

\end{document}